 \documentclass[aps,prl,preprint,groupedaddress]{revtex4}

\usepackage{graphics}
\usepackage{rotating}

\begin{document}

\title{Decay of kaonium in a chiral approach}

\author{S. P. Klevansky$^{\dag}$}
\author{R. H. Lemmer$^{\ddag}$}

\affiliation{Institut f\"ur Theoretische Physik, Universit\"at Heidelberg,
Philosophenweg 19, 69120 Heidelberg, Germany.}

\date{\today}

\begin{abstract}

The decay of the $K^+K^- $hadronic atom kaonium is investigated non--perturbatively
using meson--meson interaction  amplitudes taken from leading order
chiral perturbation theory in an approach adapted from that proposed by Oller and Oset \cite{JAO97}.
The Kudryavtsev--Popov eigenvalue equation is solved numerically for the energy shift and
decay width due to strong interactions in the $1s$ state.
These calculations introduce a cutoff $\sim 1.4$ GeV in $O(4)$ momentum space
that is necessary to regulate divergent loop contributions to the
meson--meson scattering amplitudes in the strong--interaction sector.
One finds lifetimes of $(2.2\pm 0.9)\times 10^{-18}{\rm s}$  for the ground
state of kaonium.\\


\vspace{10cm}
${}^{\dag}$Corresponding author: S. P. Klevansky

Electronic address: spk@physik.uni-heidelberg.de

$^{\ddag}$ Permanent address: School of Physics,
 University of the Witwatersrand, Johannesburg,\\  \indent Private Bag 3,
 WITS 2050, South Africa.

Electronic address: rh\_lemmer@mweb.co.za

\end{abstract}
\maketitle

 \newpage

 \subsection{1. Introduction}
   As is the case for the hadronic atoms pionium \cite{GL} and kaonic hydrogen \cite{UGM04} the energy shifts and 
   decay widths  in the coulomb spectrum of the hadronic atom $K^+K^-$, or
   kaonium \cite{AMG93}, also depend 
   primarily on the scattering length of the interacting particles, in this case
   $K^+$ and $K^-$. This scattering length, in turn, is determined by the properties of the $K\bar K$ strong
   interactions and the associated models for the possible structures of the
   $f_0(980)$  and $a_0(980)$ scalar mesons. These
   proposals include a kaon--antikaon molecular bound state
   \cite{WIsgur82,TB85,SU390,JAO03,KLS04,ZANG06,TBR08,RHL09}, a $q^2\bar q^2$
   state \cite{NNA89}, or a $q\bar q$ state \cite{LM00,NATR96,VD96,RDS98,MDS04}.

   The molecular state option has been explored in some detail in
   \cite{SU390,KLS04,ZANG06,TBR08,RHL09}. These calculations all employ a
   $SU_V(3)\times SU_A(3)$ symmetric model Lagrangian density \cite{SU390} to generate
   meson--meson interaction vertices via vector meson exchange.
   A rather different route  was followed by  Oller and Oset \cite{JAO97}. These authors
   take the  meson--meson
   interaction vertices from leading order chiral perturbation theory $(\chi$PT)  as being
   an appropriate theoretical realization of low energy QCD \cite{GL84},
   and solve the resulting  Lippmann--Schwinger equation for the corresponding $T$-matrix elements to provide a non--perturbative
  description of meson--meson scattering and reactions. This approach introduces one additional
  parameter insofar as the  meson loops occurring in
  the resulting  equations  need to be regularized with
  a high--momentum cutoff.                     
   Using a  three--momentum cutoff of $\sim 1$ GeV they find good agreement with the measured
  $\pi\pi\to \pi\pi$ phase shifts and $\pi\pi\to K\bar K$  inelasticities
   up to center-of-mass (c.m.) collision energies of order
   $\sqrt s\sim 1.2$ GeV. The masses and decay widths of the
  $f_0(980)$ and $a_0(980)$ scalar mesons are also satisfactorily reproduced as complex poles
  of the relevant $T$ matrices.

  In the following we adapt the Oller--Oset \cite{JAO97} approach to study the 
  energy shifts and decay widths introduced by the strong interactions into the
  spectrum of kaonium. 

 \subsection{2. Non--perturbative $\chi$PT scattering amplitudes}

  The calculations reported in \cite{JAO97} start from the $\chi$PT leading order interaction
  Lagrangian density \cite{JAO97,GL85,GE95} ${\cal L}_2$ for the pseudoscalar meson octet at low energies 
to generate tree--level chirally symmetric meson--meson interaction amplitudes, or 4--point vertices,
of good isospin $I$ for $s$-wave meson scattering. The explicit form of ${\cal L}_2$ is given
by Eq.~(\ref{e:La2}) of the
Appendix.
These  vertices form a symmetric matrix $V^I_{ij}$, where the labels $i$ and $j$ each refer to the channels $\{K\bar K\}$, 
$\{\pi\pi\}$ or $\{\pi^0\eta\}$. 
(We retain the labelling convention of \cite{JAO97} by setting $(i,j)=(1,2)$ where
$1$ refers to $\{K\bar K\}$ for both $I=0$ and $1$, while $2$ refers to $\{\pi\pi\}$
for $I=0$ or $\{\pi^0\eta\}$ for $I=1$).

 Knowing the $V^I_{ij}$ one can construct the coupled integral equations for the associated
 scattering matrices $T^I_{ij}$. However, for $s$--wave scattering an important
 simplification occurs. As argued in \cite{JAO97}, one can then replace
 the $V^I_{ij}$, which are in general off--shell, by their on--shell values which only depend on the square of the
 total c.m. energy $s=P_0^2$, provided that at the same time
 one uses the physical values of the pion decay constant $f\to f_\pi\approx 93$ MeV and the meson masses
 $(m_\pi,m_K, m_\eta)\approx (140,496,547)$ MeV
 that enter into the calculations. Then the coupled integral equations 
 for $T^I_{ij}$ are replaced by algebraic equations that can be solved analytically.
 In particular, one obtains $T^I_{11}(s)$ for $K\bar K$ in both isospin channels by first ``dressing '' the
 bare $K\bar K$ interaction $V^I_{11}(s)$ with $\pi\pi$ or $\pi^0\eta$ polarization loops
$\Pi^I_{22}(s)=\Pi_{\pi\pi}(s)$ or $\Pi_{\pi^0\eta}(s)$ for $I=0$ or $1$ respectively,
 and then inserting this dressed interaction into the Lippmann--Schwinger equation for
 $T^I_{11}(s)$ to find the
 result already quoted in \cite{JAO97}, 
\begin{equation}
 T^I_{11}(s) = \frac{(1-V^I_{22}\Pi^I_{22})V^I_{11}+V^I_{12}\Pi^I_{22}V^I_{21}}
{(1-V^I_{11}\Pi^I_{11})(1-V^I_{22}\Pi^I_{22})-V^I_{12}\Pi^I_{22}V^I_{21}\Pi^I_{11}}
\label{e:T11}
\end{equation}
 Here $\Pi^I_{11}(s)=\Pi_{K\bar K}(s)$ accounts for the common  $K\bar K$ polarization loop in both
 isospin channels.
 
 We characterize the various meson  loop diagrams appearing in
 $T^I_{11}(s)$ by the common symbol $-i\Pi(s)$ where
\begin{eqnarray}
 \Pi(s)=-i\epsilon\int\frac{d^4l}{(2\pi^4)}\frac{1}{(l^2-m^2_a)}\frac{1}{(l+P_0)^2-m^2_b},\quad s=P_0^2
\label{e:Pi}  
\end{eqnarray}
 that involves the integral over the two meson propagators of masses $(m_a,m_b)$ in the loop;
 $\epsilon$ is a symmetry factor \cite{JLP71} that takes on the values $(1/2,1)$  depending on whether $m_a=m_b$ refer to
 identical mesons or not. Since this integral diverges for large $4$-momenta, we introduce
 a cut--off $\Lambda$ in $O(4)$ momentum space.
 The on--shell versions of the 
 $V^I_{ij}(s)$ together with the regulated, closed form expressions for $\Pi(s)$ from
 which $\Pi_{\pi\pi}(s)$, $\Pi_{\pi^0\eta}(s)$ and
 $\Pi_{K\bar K}(s)$ may be constructed,
 are listed as Eqs.~(\ref{e:vertex}), (\ref{e:Piclosed}) and (\ref{e:eqlm}) in the Appendix.

 The poles of $T^I_{11}(s)$  on the appropriate  sheet of the cut complex $s$ plane
 are identified \cite{JAO97}
 with the masses and half widths of the scalars $f_0(600)$ (or $\sigma$) and $f_0(980)$ for $I=0$, and  $a_0(980)$ for $I=1$.
 This sheet structure  depends in turn on the analytic behaviour of the functions
$\Pi(s)$ in the denominator of Eq.~(\ref{e:T11})
that develop unitarity cuts along the real $s$-axis, starting at
 the production thresholds of $\pi\pi$, $\pi^0\eta$ and $K\bar K$ mesons at $s=(m_a+m_b)^2=4m^2_\pi$,
 $(m_{\pi^0}+m_\eta)^2$ or $4m^2_K$ respectively.
  The prescription for analytically continuing     
  $T^I_{11}(s)$  through these cuts onto the relevant  Riemann sheet in the lower half plane 
  where these poles lie is detailed in \cite{JAO97}.              

\begin{figure}
\rotatebox{0}{\includegraphics{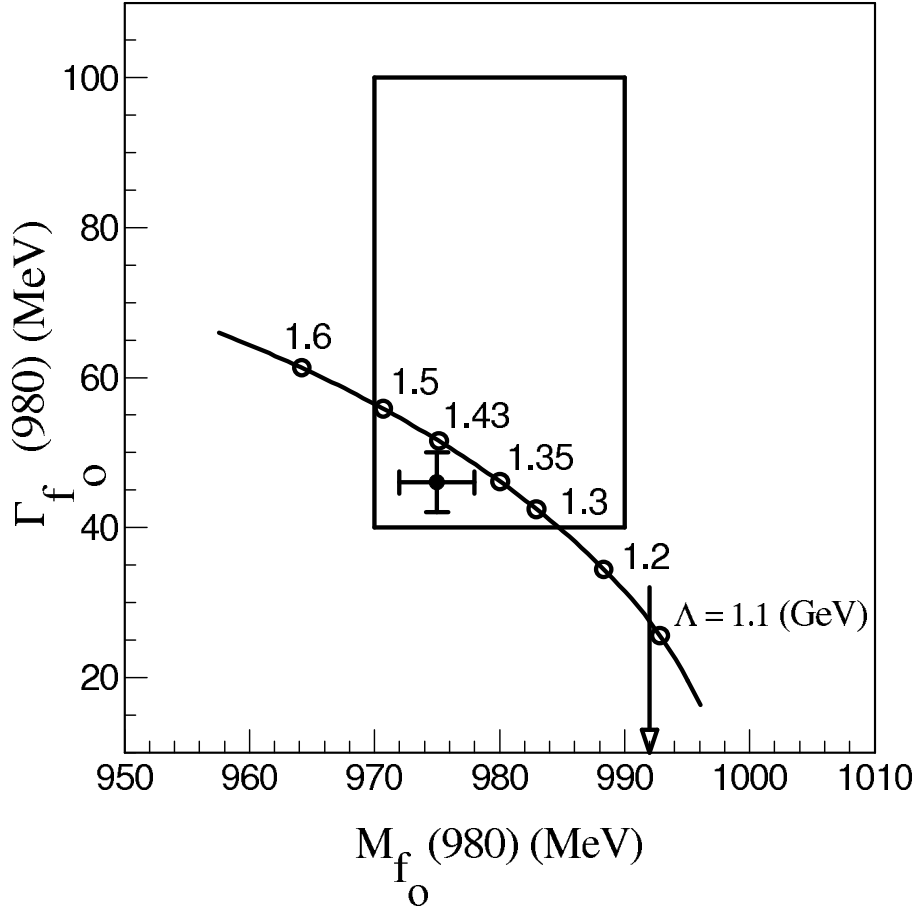}}
\caption{\label{f:fig1}                 
  Mass--width relation for $f_0(980)$ parametrized by the $O(4)$ cutoff $\Lambda$.
 The rectangle gives the limits on $M_{f_0}$ and $\Gamma_{f_0}$ suggested in the PDG data listings \cite{PDG08}, while the
 cross indicates the Fermilab E791 measurements \cite{Fermilab} and the associated error bars.
 The $K\bar K$ threshold at $2m_K\approx 992$ MeV is indicated by the arrow.
 The physical parameters used in this evaluation are $(m_\pi,m_K,f_\pi) =(140,496,93)$
 in MeV.}
 \end{figure}

    Using this information together with the closed forms for polarization functions
  and interaction vertices assembled in the Appendix, we
  reconstruct the $f_0(980)$ pole of  $T^0_{11}(s)$ as a function of
  the  cutoff in the $O(4)$ regularization scheme.
  The results are given in Fig.~\ref{f:fig1} that displays the width $\Gamma_{f_0}$ versus
  the mass $M_{f_0}$, parametrized by $\Lambda$ in order to illustrate their sensitivity to cutoff. These calculations assume
  a Breit--Wigner form $\sim (P_0-M_{f_0}+i\Gamma_{f_0}/2)^{-1}$ with
  $P_0=\sqrt{s}$ for  $T^0_{11}(s)$ in the vicinity of the pole.
  We use these results to fix $\Lambda$ by requiring that the real part of the pole coincide with
  the observed $f_0(980)$ mass as quoted in either the 
  PDG \cite{PDG08} data tables, $[(980\pm 10)-i(20\; {\rm to}\; 50)]$ MeV,
  or the Fermilab E791 experiment \cite{Fermilab} that gives $[(975\pm 3)-i(22\pm 2)]$ MeV.
  Then
  \begin{eqnarray}
   M_{f_0}-\frac{i}{2}\Gamma_{f_0}= [(980\pm 10)-i(23\;^{-7}_{+5})]\;{\rm MeV}
   \quad {\rm for}\quad \Lambda=1.35_{+0.16}^{-0.19}\; {\rm GeV}
   \label{e:L1}
   \end{eqnarray}
   in the first case, while
   \begin{eqnarray}
   \quad\quad M_{f_0}-\frac{i}{2}\Gamma_{f_0}= [(975\pm 3)-i(26\mp 2)]\;{\rm  MeV}\quad {\rm for}
   \quad \Lambda=1.43\mp 0.05\;{\rm GeV}.
   \label{e:L2}
   \end{eqnarray}
   in the second.  These choices for $\Lambda$ also reproduce the
   observed $f_0(980)$ half--widths rather satisfactorily.
   Taken together, this
  suggests that the appropriate values of the cutoff $\Lambda$ in the $O(4)$ regularization scheme
  should lie within a relatively small window around $1.4$ GeV.

  \subsection{3. Scattering lengths}
  
  The associated $K\bar K$ scattering lengths of good isospin are   
  given in terms of the $T^I_{11}(s)$ by the standard expression \cite{JLP71}
  \begin{eqnarray}
  a^I_{K\bar K}=-\lim_{s\to 4m^2_K}\frac{T^I_{11}(s)}{8\pi\sqrt{s}}=
  -\frac{T^I_{11}(4m^2_K)}{16\pi m_K}
  \label{e:aKK}
  \end{eqnarray}
  The scattering lengths $a^I_{K\bar K}$ are complex numbers with $Im\;a^I_{K\bar K}<0$  because of the presence of the open
   $\pi\pi$ and $\pi^0\eta$ decay channels for $I=0$ and $I=1$. One finds, in units of
   the inverse kaon mass
   $m^{-1}_K$, that
  \begin{eqnarray}
   a^0_{K\bar K}= 3.530-2.077i,\quad a^1_{K\bar K}= 1.651-1.416i, \quad \Lambda=1.35\;{\rm GeV}
  \label{e:a0}
  \end{eqnarray}
   or
  \begin{eqnarray}
   a^0_{K\bar K}=3.303-1.713i,\quad a^1_{K\bar K}=1.655-1.221i,
  \quad \Lambda=1.43\;{\rm GeV}
  \label{e:a1}
  \end{eqnarray}
  for the two $\Lambda$'s found in Eqs.~(\ref{e:L1}) and (\ref{e:L2}).
   
   The values calculated above for the isoscalar scattering length are similar
   in order of magnitude and sign to
    early direct experimental measurements \cite{WWE76} that yield $a^0_{K\bar K}= \big[(3.13\pm  0.30)
   -(0.67\pm 0.07)i\big]m^{-1}_K$, and a later analysis \cite{RKLL95} of $\pi\pi$ data that
   gives $a^0_{K\bar K}=(4.36-1.49i)m^{-1}_K$.
  No similar  measurements are  available for $a^1_{K\bar K}$. A model--dependent
  estimate \cite{RHL06} extracted from more recent $pp\to dK^+\bar K^0$ data \cite{KLE03} yields
   $a^1_{K\bar K}\approx [-(0.05\pm 0.05)-i(1.59\pm 0.60)] m^{-1}_K$.
   These numbers may also be compared with the rough estimate $a^0_{K\bar K}\sim
   a^1_{K\bar K}\sim (2.98-2.13i)m^{-1}_K$ from the zero--range universal approximation \cite{EBMK04}
   $M-\frac{i}{2}\Gamma\approx 2m_K-m^{-1}_K(a^I_{K\bar K})^{-2}$ for common  meson masses
   and half--widths $\sim (980-35i)$ MeV in both isospin channels.

  Moving from $T$-matrices labelled by good isospin to particle--antiparticle labels
  with the aid of Eq.~(\ref{e:KK}), one obtains the common strong interaction scattering lengths for
  the physical $K^+K^-$ and $K^0\bar K^0$ channels as
  \begin{eqnarray}
  a_{K^+K^-}=a_{K^0\bar K^0}=\frac{1}{2}\Big(a^0_{K\bar K}+a^1_{K\bar K}\Big)=
  2.591-1.747i \quad {\rm or}\quad 2.479-1.467i
  \label{e:a+-} 
  \end{eqnarray}
   for cutoffs $(1.35,\;1.43)$ GeV respectively, if the  $K^0K^+$ mass 
   difference $\Delta = m_{K^0}-m_{K\pm}\approx 4$ MeV is ignored. If not, then
 the $K^+K^-$ scattering length becomes \cite{ADM70} 
\begin{eqnarray}                                                
a_p=\Big(\frac{a_{K^+K^-}-k_0a^0_{K\bar K }a^1_{K\bar K}}{1-k_0a_{K^+K^-}}\Big)
=2.688-1.896i\quad {\rm or}\quad 2.567-1.566i
\label{e:AP} 
\end{eqnarray}
 for the two cutoffs in question. Here $k_0=\sqrt{2m_{K^0}\Delta}\;.$

\subsection{4. Strong interaction energy shifts and decay widths of kaonium}

The level shifts and decay widths for kaonium due to strong interactions have
been discussed in detail in \cite{KLS04,RHL09} using the vector meson exchange
 model Lagrangian density \cite{SU390} mentioned in the Introduction.
We re--evaluate these quantities using the basic $\chi$PT ${\cal L}_2$ density 
given by Eq.~(\ref{e:La2}) instead.

The unshifted ground state of kaonium
lies at $E_{1s}=-\frac{1}{2}\alpha^2\mu=-6.576$ keV where $\mu=\frac{1}{2}m_{K^{\pm}}$ is the reduced mass with $m_{K^{\pm}}=494$  MeV
and $\alpha\approx 1/137$.
Contrary to the case of pionium, however,
where charge exchange  $\pi^+\pi^-\to \pi^0\pi^0$ dominates the decay \cite{GL}, $K^+K^-\to K^0\bar K^0$ is not allowed
due the $K^0K^{\pm}$ mass difference. Thus the principal strong decay modes are kaonium $\to \pi\pi +\pi^0\eta$
that proceed via strange quark annihilation.

Isospin is broken by the coulomb field as well as the  meson mass difference.
Since the binding energy of kaonium is small relative to that of the
strongly interacting $K\bar K$ ground state of $f_0(980)$ at $\sim 10$ MeV, the modified energy spectrum can be found by the standard procedure \cite{Bethe49}
of joining  the zero momentum $s$-wave scattering function of mixed isospin $u(r)=r\psi(r)\to 1-r/a_p$
of the pair emerging 
from the strong interaction zone with scattering length $a_p$, onto an
exponentially decaying pure coulomb wave at infinity. This is given by an
incoming Whittaker function \cite{AS} $ W_{i\eta,1/2}(2ikr)$ at complex momentum $k=k_\lambda=-i\mu\alpha\lambda$
with ${\rm Re}\lambda>0$. The coulomb parameter index for attractive fields then reads $i\eta=-i\mu\alpha/k=1/\lambda$. Here $\lambda=\lambda_n$,
$n=1,2,\cdots$ is a set of eigenvalues to be determined by the logarithmic matching condition at $r=d$ outside the strong interaction zone of
extent $\sim 1/m_K$.
One thus retrieves the Kudryavtsev--Popov eigenvalue equation \cite{KP79} which we rearrange
as       
\begin{eqnarray}
\alpha_{pc}=
2\mu\alpha [\frac{\lambda_n}{2}+\ln\lambda_n+\psi(1-\frac{1}{\lambda_n})+\gamma],\quad n=1,2,\cdots
\label{e:KP}
\end{eqnarray}
  Here $\psi$ is the
 standard digamma function \cite{AS}, $\gamma=0.57721\cdots$ the Euler constant and
 $\alpha_{pc}=1/a_{pc}$ is the physical $K^+K^-$ inverse scattering length in the presence coulomb interactions as defined
by Bethe \cite{Bethe49},
 \begin{eqnarray}
  \alpha_{pc}=\alpha_p-2\mu\alpha[\ln(2\mu\alpha d)+\gamma]
  \label{e:PC}
  \end{eqnarray}
 with $\alpha_p=1/a_p$ from Eq.~(\ref{e:AP}). The role of $\alpha_{pc}$ as the
 relevant experimental observable has also been stressed in \cite{BH99}.
We estimate $d\approx 2.2m^{-1}_K$ for the  (calculable) Bethe joining radius \cite {Bethe49} for the coulomb plus strong interaction field; $d$
is thus fixed and not a parameter. Since $d$ is much smaller than the Bohr radius $1/\mu\alpha$ of kaonium, the small argument approximation for $W_{i\eta,1/2}(2ikr)$
suffices for deriving Eq.~(\ref{e:KP}).

The revised $s$-wave  energy levels and decay widths ($\lambda_n$ is complex because
$\alpha_p$ and therefore $\alpha_{pc}$ is complex) of the kaonium atom are given by
 \begin{eqnarray}                                       
 E_{\lambda_n}-\frac{i}{2}\Gamma_{\lambda_n}=\frac{(k_{\lambda_n})^2}{2\mu}=-\frac{1}{2}\alpha^2\lambda_n^2\mu
 \label{e:shiftwidth}
 \end{eqnarray}
   In the absence of any strong interactions $a_{p}\to 0$ so that $\alpha_{pc}\to\infty$.
   From Eq.~(\ref{e:KP})
   this means that $\lambda_n$ is then determined by the poles of $\psi(1-\frac{1}{\lambda_n})$
   as $\lambda_n^{-1}= n$, and one recovers the pure coulomb spectrum from Eq.~(\ref{e:shiftwidth}).

   The input for $\alpha_{pc}$ on the left hand side of Eq.~(\ref{e:KP}) is obtained
   from Eqs.~(\ref{e:AP}) and (\ref{e:PC}) as
   \begin{eqnarray}
   \alpha_{pc}=(2.589-1.654i)^{-1}m_K\quad {\rm or}\quad (2.459- 1.375i)^{-1}m_K
   \label{e:pcnum}
   \end{eqnarray}
  for $\Lambda=1.35$ or $1.43$ GeV respectively.  Eq.~(\ref{e:KP}) has multiple roots. Solving for the
  eigenvalue $\lambda_1$ corresponding to the $1s$ ground state of kaonium, one finds
  $\lambda_1= 0.9812+0.0118i$ or $\lambda_1=0.9822+0098i$ for the above two values of the cutoff.
  The energy shifts $\Delta E_{1s}=(E_{\lambda_1}-E_{1s})$,
  half--widths $\frac{1}{2}\Gamma_{1s}=\frac{1}{2}\Gamma_{\lambda_1}$ and resulting
  lifetimes $\tau_{1s}=\hbar/\Gamma_{1s}$ (in units $10^{-18}{\rm s}$)
  introduced by the strong  interactions
  are then \footnote{It is interesting to note that these shifts and widths, which have been generated using chiral interaction amplitudes, are
  bracketed by those based on local \cite{KLS04} or non--local \cite{RHL09}
  model potentials to describe the strong interactions.}
  \newpage
  \begin{eqnarray}
  \Delta E_{1s}-\frac{i}{2}\Gamma_{1s}
  &=& [(246^{+21}_{-25})-\frac{i}{2}(304^{+194}_{-88})]\:{\rm eV},\quad\tau_{1s}=2.2\mp 0.9,
 \quad \Lambda=1.35^{-0.19}_{+0.16}\:{\rm GeV}
  \nonumber\\                       
  &=&[(233^{+8}_{-7})- \frac{i}{2}(254^{+28}_{-24})]\:{\rm eV},
 \quad\tau_{1s}=2.6\mp 0.3,\quad\Lambda=1.43\mp 0.05\:{\rm GeV}
  \nonumber\\
 \label{e:delEG}
 \end{eqnarray}
  Notice that the energy shifts are repulsive and of the same order of magnitude as the decay
  widths.

  The uncertainties on $\Lambda$  have been also
  included in these calculations. The larger error bars at the lower cutoff clearly
  overlap with those at the upper cutoff. A conservative estimate
  for the lifetime is thus provided by the spread of values at the lower cutoff of $\tau_{1s}\sim 1$ to $3\times 10^{-18}{\rm s}$,
  rounded to the nearest integer.

\begin{figure}
\rotatebox{0}{\includegraphics{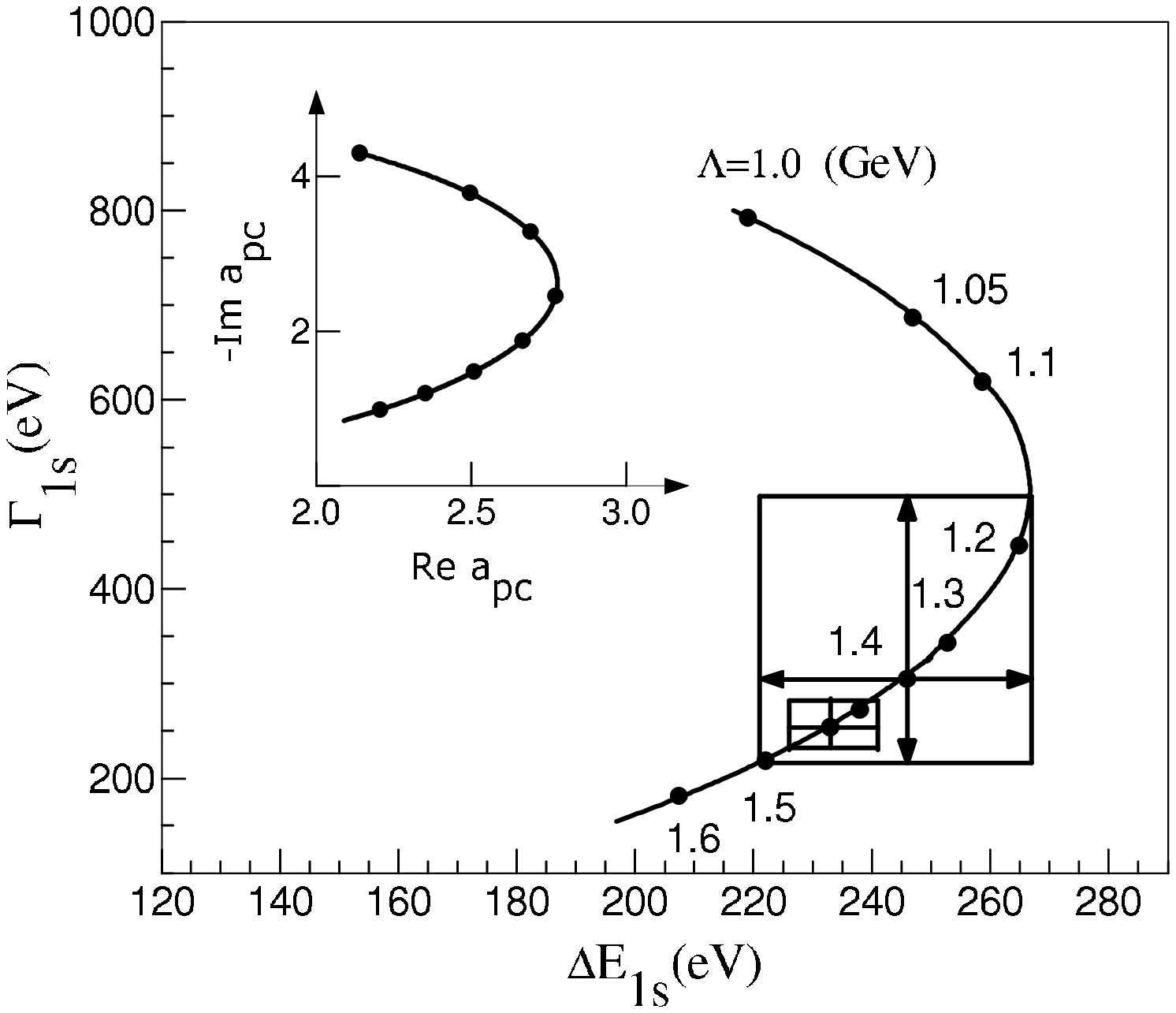}}
\caption{\label{f:fig2}
The $1s$ decay width $\Gamma_{1s}$ versus energy shift $\Delta E_{1s}$ for kaonium
parametrized by the $O(4)$ cutoff $\Lambda$.
The estimated energy shift and corresponding width can lie anywhere along the arc of the curve
intersecting the calculated uncertainties--on--$\Lambda$ rectangles centered at $\Lambda=1.35$ or $1.43$ GeV
(large and small rectangles) respectively.
These rectangles correspond to fitting either the PDG \cite{PDG08} or the  Fermilab E791 data \cite{Fermilab} for the observed
$f_0(980)$ mass within the experimental uncertainties, refer Fig.~\ref{f:fig1}.
The inset shows the
$-Im\; a_{pc}$ versus $Re\; a_{pc}$ parametric curve in units $m^{-1}_K$ calculated for the same set of $\Lambda'$s.
The shape of this curve mimics that of the width--shift curve. The latter also turns in the close vicinity of $\Lambda =1.18$ GeV where the parametric
$a_{pc}$ curve turns as $Re\;a_{pc}$ reaches
its maximum value in the inset.}                                             
\end{figure}

  Fig.~\ref{f:fig2}  gives a parametric plot  of $\Gamma_{1s}$ versus $\Delta E_{1s}$ for a range
 of $\Lambda'$s. The inset in this figure shows an analogous plot of the quantities $-Im\;a_{pc}$ versus $Re\;a_{pc}$
 that provide the input for generating the main curve.
 One sees  that the two curves behave  in a similar fashion,
 and moreover turn at approximately the same value  $\Lambda=1.18$ GeV where $Re\;a_{pc}$ reaches
 its maximum value in the inset of Fig.~\ref{f:fig2}.  
 This behavior is confirmed by the original
 Deser ${\it et\;al.}$ approximate formula \cite{SD54}
 that is recovered by expanding the roots of Eq.~(\ref{e:KP}) to lowest order in the interaction parameter $2\mu\alpha a_{pc}$
 about their limiting values $\lambda_n^{-1}=n$. Then 
 
 \begin{eqnarray}
 \Delta E_{1s}-\frac{i}{2}\Gamma_{1s} \approx 2\mu^2 \alpha^3a_{pc}
 =\frac{2\pi}{\mu}a_{pc}|\psi_{1s}(0)|^2
 \label{e:Deser}
 \end{eqnarray}
 for the $1s$ ground state, where $\psi_{1s}(0)=(\mu\alpha)^{3/2}\pi^{-1/2}$.
 In this limit one sees that the two parametric curves in Fig.~\ref{f:fig2} are related by a simple rescaling of
 their axes.

     The Deser estimate, Eq.~(\ref{e:Deser}) gives 
  \begin{eqnarray}
  \Delta E_{1s}-\frac{i}{2}\Gamma_{1s}
  &\approx& [(248^{+19}_{-19})-\frac{i}{2}(318^{+202}_{-94})]\:{\rm eV},\quad\tau_{1s}=2.1\mp 0.9,
 \quad \Lambda=1.35^{-0.19}_{+0.16}\:{\rm GeV}                                            
  \nonumber\\                       
  &\approx&[(236^{+7}_{-7})- \frac{i}{2}(264^{+30}_{-26})]\:{\rm eV},
 \quad\tau_{1s}=2.5\mp 0.3,\quad\Lambda=1.43\mp 0.05\:{\rm GeV}
  \nonumber\\
 \label{e:delDES}
 \end{eqnarray}
 for the energy shift and width in the place of Eq.~(\ref{e:delEG}).
 In the present instance this approximation is seen to be fairly reliable,  overestimating the
 energy shifts and widths  by $\sim 1\%$ and $\sim 5\%$ respectively, when compared with the eigenvalue solutions
 obtained from the Kudryavtsev--Popov  equation. 

  \subsection{5. Summary and conclusions}

  We have studied the  energy shift and decay width of the $1s$ ground state  of
  kaonium  using interactions taken from chiral perturbation
  theory to construct the strong $K\bar K$ scattering amplitudes. These calculations contain
  a single regulating cutoff $\Lambda$
   that is fixed  by requiring that the relevant pole of the $K\bar K$  scattering amplitude in the
  isoscalar channel reproduce the experimental $f_0(980)$ mass determinations within their quoted
  error bars \cite{PDG08,Fermilab}.

  The calculated decay lifetimes for kaonium then range from $\sim 1$ to $3\times 10^{-18}{\rm s}$.
  Direct experimental information on the energy shift and decay width of kaonium
  is not  yet available.
  However, such experiments
   have already been performed in the case of
  kaonic hydrogen ($K^-p$) by the DEAR collaboration \cite{DEAR05}.
 These data have been analyzed using a variety of theoretical approaches to extract scattering lengths,
 see for example \cite{UGM04,BNW05,BUN06}.

A similar experimental program for kaonium would allow one to extract a value
for the physical $K^+K^-$  scattering length $a_{pc}$ by combining Eqs.~(\ref{e:shiftwidth})  and
(\ref{e:KP}) to determine a value for $\lambda_1$ and hence $\alpha_{pc}$.
This information would also allow  a re--assessment of the various theoretical models
mentioned in the Introduction for the  $f_0(980)$ and $a_0(980)$ scalar mesons
by comparing with
their predictions for the strong scattering lengths $a^I_{K\bar K}$ and hence $a_{pc}$.
The additional feature \cite{HPO82} that kaonium 
is the only purely mesonic atom with hidden strangeness that decays 
via strangeness annihilation makes such  experimental and theoretical investigations
of particular interest.

Indirect methods  of observation of the formation and decay of kaonium \cite{AMG93,BK95,SVB96} may also be feasible in the future.

 \subsection{Acknowledgments}
 One of us (RHL) would like to thank the Ernest Oppenheimer Memorial Trust
 for research support in the form of  a Harry Oppenheimer Fellowship. The kind hospitality
 of the Institut f\"ur Theoretische Physik, Universit\"at Heidelberg, is also gratefully acknowledged.
 We would also like to thank A. Gal for bringing the second reference included
 in \cite{ADM70} to our attention.

\subsection{6. Appendix}
 \subsubsection{(i) Lagrangian density ${\cal L}_2$ and the $4$--point interaction vertices}
  The leading order $\chi$PT interaction
  Lagrangian density for the pseudoscalar meson octet  
  used in \cite{JAO97} is
\begin{eqnarray}
{\cal L}_2=\frac{1}{12f^2}Tr\Big((\partial_\mu\Phi\Phi-\Phi\partial_\mu\Phi)^2+M\Phi^4\Big)
\label{e:La2}
\end{eqnarray}
where
\[\\
\Phi
 \\
=\left(\begin{array}{ccc}
\frac{1}{\sqrt 2}\pi^0+\frac{1}{\sqrt 6}\eta&\pi^+&K^+\\
\pi^-&-\frac{1}{\sqrt 2}\pi^0+\frac{1}{\sqrt 6}\eta&K^0\\
 K^-&\bar K^0&-\frac{2}{\sqrt 6}\eta\end{array}\right )
\]
is a $3\times 3$ matrix of the meson fields in $SU(3)$ flavor space constructed from the Gell-Mann $\lambda$ matrices that are the generators
of this group; $\eta$ is identified with $\eta_8$.
$M$ is the diagonal mass matrix $M= \text{diag}(m^2_\pi,m^2_\pi, 2m^2_K-m^2_\pi)$ and $f$ is the pion decay constant.
The trace runs over the $SU(3)$ flavor
space.
 
 Call $iV^I_{ij}(s)$  the set of $4$-point vertex diagrams generated by $i{\cal L}_2$.
 Working in a set of basis states of good isospin
 \begin{eqnarray}
 |(K\bar K)^{0,1}>=-\frac{1}{\sqrt 2}[K^+K^-\pm K^0\bar K^0]
\label{e:KK}
 \end{eqnarray}
 and
 \begin{eqnarray}
  &&|(\pi\pi)^0>= -\frac{1}{\sqrt 3}[ \pi^+\pi^-+\pi^-\pi^++\pi^0\pi^0],
  \quad |(\pi^0\eta)^1>=\pi^0\eta
  \label{e:pipi}
  \end{eqnarray}
   the on--shell values   
  \footnote{Note that since the
   pions act as identical bosons in this basis, some authors e.g.
   \cite{SU390,JAO97},
   include an additional normalization of $1/\sqrt 2$ in the definition of
   $|(\pi\pi)^0>$. Hence the matrix elements $V^0_{ij}$ given in Eq.~(\ref{e:vertex}) are larger
   by a factor $\sqrt{2}$ than   
   those listed by Oller and Oset for each pion label 2 appearing on the matrix.}
   for the $V^I_{ij}$ are
  \begin{eqnarray}
 &&V^0_{11}= \frac{3}{4}\frac{s}{f^2},\quad  V^0_{21}=\frac{1}{2}\sqrt{\frac{3}{2}}
  \frac{s}{f^2},\quad V^0_{22} =\frac{1}{f^2}(2s-m_\pi^2)
  \nonumber
  \\
  &&V^1_{11}= \frac{1}{4}\frac{s}{f^2},\quad V^1_{21} =-\sqrt{\frac{2}{3}}\frac{1}{f^2}
  (\frac{3}{4}s-\frac{1}{12}m^2_\pi-\frac{1}{4}m^2_\eta -\frac{2}{3}m^2_K),\quad
  V^1_{22}= \frac{1}{3}\frac{m^2_\pi}{f^2}
  \label{e:vertex}
  \end{eqnarray}
   Here $\sqrt s$ is the total
   collisional energy  in the CM system.

 \subsubsection{(ii) Polarization loop integrals}

 The expression for the $O(4)$ regularized integral in Eq.~(\ref{e:Pi})
 depends on where $s$ lies relative to the cut that starts at the branch point
 $(m_a+m_b)^2$. For  $s>(m_a+m_b)^2$ on the upper lip of the cut along the real axis,
 the integral acquires an imaginary part and one finds
 \begin{eqnarray}
\Pi(s)=\frac{\epsilon}{(4\pi)^2}\Big[\frac{m^2_a}{m^2_a-m^2_b}
\ln(1+\frac{\Lambda^2}{m^2_a})-\frac{m^2_b}{m^2_a-m^2_b}\ln(1+\frac{\Lambda^2}{m^2_b})
-L_{ab}(s)\Big]
\label{e:Piclosed} 
\end{eqnarray}
 with $L_{ab}(s)$  given by 
\begin{eqnarray}
&&L_{ab}(s)=-1-\frac{1}{2}\Big(\frac{m^2_a+m^2_b}{m^2_a-m^2_b}-\frac{ m^2_a-m^2_b}{s}\Big)
\ln\frac{m^2_a}{m^2_b} 
\nonumber\\
&&+\sqrt{f_{ab}}\Big[\tanh^{-1}(\frac{\sqrt{f_{ab}}}{1-\frac{m^2_a-m^2_b}{s}})
 +\tanh^{-1}(\frac{\sqrt{f_{ab}}}{1+\frac{m^2_a-m^2_b}{s}})\Big]-i\pi\sqrt{f_{ab}},
 \quad s>(m_a+m_b)^2
\nonumber\\                                      
&&\sqrt{f_{ab}} =\Big(1-\frac{(m_a-m_b)^2}{s}\Big)^{1/2}
\Big(1-\frac{(m_a+m_b)^2}{s}\Big)^{1/2} =\frac{2p_{ab}}{\sqrt s}
\label{e:Lab}
\end{eqnarray}
 and $p_{ab}$ is the magnitude of the 3-momentum of either meson in the c.m. system.
 Note that all expressions are symmetric under the interchange $m_a\to m_b$.
Equation (\ref{e:Piclosed}) with $m_a=m_\pi$ and $m_b=m_\eta$ gives the closed form for $\Pi_{\pi^0\eta}(s)$.

 For equal masses $m_a=m_b=m$, $\Pi(s)$ reduces to the simple form 
\begin{eqnarray}                                                  
&&\Pi(s)=\frac{\epsilon}{(4\pi)^2}\Big[1+\ln(1+\frac{\Lambda^2}{m^2})+\frac{m^2}{\Lambda^2}
(1+\frac{m^2}{\Lambda^2})^{-1}
-2\sqrt f\tanh^{-1}\sqrt f +i\pi\sqrt f\Big],\quad s>4m^2
 \nonumber\\
 &&\sqrt f=\Big(1-\frac{4m^2}{s}\Big)^{1/2}
 \label{e:eqlm}
\end{eqnarray}
 Setting $m=m_\pi$ or $m_K$ with $\epsilon=1/2$ or $1$ respectively then
 leads to closed forms for $\Pi_{\pi\pi}(s)$ and  $\Pi_{K\bar K}(s)$.


\newpage

\newpage             


\subsection{References}



\noindent

                                       


\begin{thebibliography}{17}
\bibitem{GL}

For recent reviews, see J. Gasser, V.E. Lyubovitskij, and A. Rusetsky,
  Phys. Rept. {\bf 456}, (2008) 167; Ann. Rev. Nucl. Part. Sci. {\bf 59},
 (2009) 169.
 

\bibitem{UGM04}
Ulf--G. Mei{\ss}ner, U. Raha and A. Rusetsky, Eur. Phys. J. C {\bf 35},
(2004)  349.


\bibitem{AMG93}
S. Wycech and A. M. Green, Nucl. Phys. A {\bf  562}, (1993) 446.




\bibitem{WIsgur82}
J. Weinstein and N. Isgur, Phys. Rev. Lett. {\bf 48}, (1982)  659; Phys. Rev.
D {\bf 27}, (1983) 588; D {\bf 41}, (1990)  2236.



\bibitem{TB85}

T. Barnes, Phys. Lett. B {\bf 165 }, (1985) 434.

\bibitem{SU390}
D. Lohse, J. W. Durso, K. Holinde and J. Speth, Nucl. Phys. A {\bf 516},
(1990) 513; G. Jan{\ss}en, B. C. Pearce, K. Holinde and J. Speth, Phys. Rev.
 D {\bf 52}, (1995) 2690.


\bibitem{JAO03}

J. A. Oller, Nucl. Phys. A {\bf 714}, (2003) 161.

\bibitem{KLS04}
S. Krewald, R. H. Lemmer and F. P. Sassen, Phys. Rev. D {\bf 69},
  (2004) 016003.

\bibitem{ZANG06}
Y.-J. Zhang, H.-C. Chiang, P.-N. Shen, and B.-S. Zou, Phys. Rev. D {\bf 74},
(2006) 014013.

\bibitem{TBR08}
 T. Branz, T. Gutsche and V.E. Lyubovitskij, Eur. Phys. J. A {\bf 37},
  (2008) 303.
  
\bibitem{RHL09}
R. H. Lemmer, Phys. Rev. C {\bf 80}, (2009) 045205.
                                                 

\bibitem{NNA89}

N. N. Achasov and V. N. Ivanchenko, Nucl. Phys.  B {\bf 315}, (1989)  465.

\bibitem{LM00}

L. Montanet, Nucl. Phys. B (Proc. Suppl.) {\bf 86}, (2000)  381; V.V. Anisovich
{\it et al.}, Phys. Lett. B {\bf 480}, (2000) 19.

\bibitem{NATR96}
N. A. Tornquist and M. Roos, Phys. Rev. Lett. {\bf 76}, (1996) 1575.


\bibitem{VD96}
V. Dmitrasinovi$\acute{c}$, Phys. Rev. C {\bf 53}, (1996) 1383.

\bibitem{RDS98}
R. Delbourgo and M. D. Scadron , Int. J. Mod. Phys. A {\bf 13}, (1998) 657.


\bibitem{MDS04}
M. D. Scadron, G. Rupp, F. Kleefeld, and E. van Beveren, Phys. Rev. D {\bf 69}, (2004)  014010;
Erratum, Phys. Rev. D {\bf 69}, (2004)  059901.
        



\bibitem{JAO97}
J. A. Oller and E. Oset, Nucl. Phys. A {\bf  620}, (1997) 438;
Erratum, Nucl. Phys. A {\bf  652}, (1997) 407.

\bibitem{GL84}

J. Gasser and H. Leutwyler, Ann. Phys. {\bf 158}, (1984) 142.                        


\bibitem{GL85}

J. Gasser and H. Leutwyler, Nucl. Phys. B {\bf  250}, (1985) 465.



\bibitem{GE95}
G. Ecker, Prog. Part. Nucl. Phys. {\bf 35}, (1995) 1.


\bibitem{JLP71}

J. L. Peterson, Phys. Rep. {\bf C 2}, (1971) 158.

 
\bibitem{PDG08}
 K. Nakamura et al. (Particle Data Group) JPG {\bf 37}, (2010) 075021.

\bibitem{Fermilab}
 Fermilab E791 Collaboration, E. M. Aitala {\it et al.,} Phys. Rev. Lett. {\bf
 86}, (2001)  765.



\bibitem{WWE76}
W. W. Wetzel {\it et al.}, Nucl. Phys. B {\bf 115}, (1976) 208.

\bibitem{RKLL95}

R. Kami\'{n}ski and L. Le\'{s}niak, Phys. Rev. C {\bf 51}, (1995) 2264.

\bibitem{RHL06}
R. H. Lemmer, Phys. Lett. B {\bf 633}, (2006) 265.

\bibitem{KLE03}
V. Kleber, {\it et al.}, Phys. Rev. Lett. {\bf 91}, (2003) 172304.


\bibitem{EBMK04}
E. Braaten and M. Kusunoki, Phys. Rev. D {\bf  69} (2004) 074005;
M. B. Voloshin, Phys. Lett. B {\bf  579} (2004) 316.



\bibitem{ADM70}
 A. D. Martin and G. G Ross, Nucl. Phys. B {\bf 16}, (1970)  479;
 R. H. Dalitz and S. F. Tuan, Ann. Phys. {\bf 8} (1959) 100.



 \bibitem{Bethe49}
H. A. Bethe, Phys. Rev. {\bf 76}, (1949) 38; J. D. Jackson and J. M. Blatt,
Rev. Mod. Phys. {\bf 22}, (1950) 77.





\bibitem{AS}
 M. Abramowitz and I. A. Stegun, Editors, {\it Handbook of Mathematical
 Functions} (Dover Publications, Inc., New York, 1965).

 
\bibitem{KP79}

A.E. Kudryavtsev and V. S. Popov, JETP Lett. {\bf 29}, 280 (1979); V. S. Popov,
A.E. Kudryavtsev, and V. D. Mur, Sov. Phys. JETP {\bf 50}, 865 (1979).

\bibitem{BH99}

B. Holstein, Phys. Rev. D {\bf 60}, (1999) 114030.


\bibitem{SD54}

S. Deser, M. L. Goldberger, K. Baumann, and W. Thirring, Phys. Rev. {\bf 96},
 (1954) 774; T. L. Trueman, Nucl. Phys. {\bf 26}, (1961) 57.

  




\bibitem{DEAR05}
 M. Cargnelli ${et.\;al}$
 [DEAR collaboration], in Proceedings of ''HadAtom'' Workshop, 13--17 October 2003,
 ECT* (Trento, Italy), arXiv:hep-ph/0401204; {\it ibid.}
 Int. J. Mod. Phys. A {\bf 20}, (2005);
G Beer {\it et al.}, [DEAR collaboration] Phys. Rev. Lett. {\bf 94}, (2005) 212302.

 \bibitem{BNW05}

 B. Borasoy, R. Ni{\ss}ler, and W. Weise, Phys. Rev. Lett. {\bf 94}, (2005) 213401.

 \bibitem{BUN06}

 B. Borasoy, U.-G. Mei{\ss}ner, and R. Ni{\ss}ler, Phys. Rev. C {\bf 74}, (2006)
 055201.
 
 \bibitem{HPO82}

 H. Poth, Invited paper, Workshop on Physics at Lear with Low--Energy Cooled Antiprotons, 9--16 May
 1982, Erice, Sicily, CERN--EP/82--82, 24 June 1982.


\bibitem{BK95}
B. Kerbikov, Z. Phys. A {\bf 353}, (1995) 113.


\bibitem{SVB96}
S. V. Bashinsky and B. Kerbikov, Phys. Atm. Nucl. {\bf 59}, (1996) 1979.



\end{thebibliography}
\end{document}